# On the Possibility of Nonlinearities and Chaos Underlying Quantum Mechanics

Wm. C. McHarris

Departments of Chemistry and Physics/Astronomy
Michigan State University
East Lansing, MI 48824, USA
E-mail: mcharris@cem.msu.edu

## Abstract

Some of the so-called imponderables and counterintuitive puzzles associated with the Copenhagen interpretation of quantum mechanics appear to have alternate, parallel explanations in terms of nonlinear dynamics and chaos. These include the mocking up of exponential decay in closed systems, possible nonlinear extensions of Bell's inequalities, spontaneous symmetry breaking and the existence of intrinsically preferred internal oscillation modes (quantization) in nonlinear systems, and perhaps even the production of diffraction-like patterns by "order in chaos." The existence of such parallel explanations leads to an empirical, quasi-experimental approach to the question of whether or not there might be fundamental nonlinearities underlying quantum mechanics. This will be contrasted with recent more theoretical approaches, in which nonlinear extensions have been proposed rather as corrections to a fundamentally linear quantum mechanics. Sources of nonlinearity, such as special relativity and the measurement process itself, will be investigated, as will possible implications of nonlinearities for entanglement and decoherence. It is conceivable that in their debates both Einstein and Bohr could have been right—for chaos provides the fundamental determinism favored by Einstein, yet for practical measurements it requires the probabilistic interpretation of the Bohr school.

# I. Introduction

Ever since the renaissance of science with Galileo and Newton, scientists—and physicists, in particular—have been unabashed reductionists. Complex problems have been disassembled into their simpler and more readily analyzable component parts, which have been further dissected and analyzed, without foreseeable end. Naturally, this has led to deeper and deeper underlying layers for explaining nature, each succeeding lower level presumably being more fundamental than the previous one. Often these layers of analysis and interpretation have led to unifications, and some scientists, primarily elementary particle physicists and cosmologists, have even been known to tout recently that they could be within striking distance of a grand unified "Theory of Everything."

Within the last several decades, progress in nonlinear dynamics, and particularly in its extreme manifestations, viz., chaos theory, has led to some questioning of this runaway reductionism. Much of nature now appears to be nonlinear, even chaotic, and it is not at all clear that every complex problem can be disassembled into non-interacting component parts. In fact, strongly nonlinear systems can behave fundamentally differently from their linear counterparts—and differently from systems in which nonlinearities are introduced as perturbations ("distorted linear problems," as termed by Feigenbaum [1]). Superficially, the behavior of strongly nonlinear systems can appear every bit as perplexing and counterintuitive as that of the Copenhagen interpretation of quantum mechanics.

There have been objections to the Copenhagen interpretation ever since the Einstein-Bohr debates of the 1930's [2-4], but partly because of the persuasiveness of Bohr, this interpretation became orthodox and held sway without too much competition throughout the twentieth century. However, with the recent advent of developments in quantum information theory and the siren beckoning of the possibilities in quantum computing and perhaps even in further-out concepts such as quantum teleportation, there has been a revival of interest in revisiting the fundamentals of quantum mechanics. For example, there are now at least two prominent on-going series of international workshops devoted to this topic, the Workshops on "Mysteries, Puzzles and Paradoxes in Quantum Mechanics" [5] held in Gargnano, Italy, and the Conferences on "Quantum Theory: Reconsideration of Foundations" [6] held in Växjö, Sweden.

Two classes of proffered alternative interpretations of quantum mechanics have involved nonlinearities and so-called "hidden variables" (although not initially apparent, there are conceptual similarities between these two classes, especially when the nonlinearities are introduced as perturbative add-ons). Among the most enduring of the former was the "theory of the double solution" of de Broglie, which was published in a number of guises over a period of several decades [7]. In this theory the standard probabilistic $\Psi$-wave of the Schrödinger equation is accompanied by a very similar $u$-wave, which contains a (necessarily nonlinear) singularity representing the particle behavior. Among the more discussed hidden variables theories was the "pilot wave" theory of Bohm [8], in which a hidden "deterministic" pilot wave guides the normal probabilistic wave. Despite the prominence of their authors, neither of these interpretations garnered much credence, and they more or less fell by the wayside, along with other, more fanciful theories such as Everett's relative state ("many universes") formulation of quantum mechanics [9].

More recently, nonlinear extensions to quantum mechanics have been proposed by Weinberg [10], Gisin [11], Mielnik [12], and Czachor and Doebner [13]. These have been mostly abstract theoretical approaches, attacking the problem "from the top down," to borrow terminology from the literature of chaos and emergent systems. They have encountered some rather severe difficulties, including the question of whether or not including nonlinearities introduces superluminal (nonphysical) signals into multi-particle correlations. Mielnik [12] sums up their current thinking succinctly:

> "I cannot help concluding that we do not know truly whether or not nonlinear QM generates superluminal signals—or perhaps, it resists embedding into too narrow a scheme of tensor products. After all, if the scalar potentials were an obligatory tool to describe the vector fields, some surprising predictions could as well arise! …the nonlinear theory would be in a peculiar situation of an Orwellian 'thoughtcrime' confined to a language in which it cannot be even expressed. …A way out, perhaps, could be a careful revision of all traditional concepts…"

Meanwhile, Bell [14] had furnished new insight into the Einstein, Podolsky, and Rosen (EPR) paradox [2]. While doing so, he demonstrated that von Neumann's proof [15] that "local hidden variables" must be excluded from a self-consistent theory of quantum mechanics was flawed. What Bell did was to insert statistical, potentially experimental meaningfulness into Bohm's more specific (spin-1/2) reduction [16] of the EPR paradox. In essence, Bell's inequality places an upper bound on the statistical correlations for measurements made by widely separated (non-communicating) experimentalists on correlated pairs of particles. Classical mechanics (or statistics) requires that this upper bound be obeyed, whereas quantum mechanics, utilizing "entangled pairs," allows it to be raised (violated). Although Bell's work was slow to attract initial attention, by the end of the 1960's it had begun to work its way toward the forefront of attention in reconsiderations of the foundations of quantum mechanics. Numerous variants of Bell's inequalities appeared [17], including the experimentalist-friendly version formulated by Clauser, Horne, Shimony, and Holt (the CHSH inequality) [18]. These were exploited during the next three decades by increasingly sophisticated experiments [17], essentially all of which demonstrated violation of Bell's inequalities. Interpretations varied, but most of them included some inference that "local reality" was excluded and Einstein's "spooky action at a distance" was indicated. Quantum mechanics—specifically the Copenhagen interpretation—triumphed. (Bell's inequalities, including some very recent work aimed at disproving their basic underlying assumptions, will be covered in some detail in Section IV below.)

Quantum mechanics has always been regarded as the quintessential linear science. As an example, the initial chapter in Dirac's classic text [19] is the "The Principle of Superposition." Additionally, quantum mechanics has been considered to be *the fundamental science*, with any sort of classical mechanics being only an approximation, producing the same results only when the quantum of action $\hbar$ approaches zero or the object becomes large enough that the de Broglie wavelength is insignificant (the Correspondence Principle).

Now, during the last several decades chaos theory has been applied with great success to a broad spectrum of scientific fields, ranging from biology to physics, from mathematics to economics, and even to traffic patterns and word analysis in literature. The most noteworthy field where it has met with only limited success is quantum chaos—although books have been written on the subject [20], a large number of chaoticists still question whether or not such a field as quantum chaos even exists [21]. Many of the applications have involved "quantum chaotic billiards" [20,21], in which quantum systems actually appear simpler than their classical counterparts; others involve items such as the presumably chaotic spacing of, say, nuclear energy levels [22], which could have other origins (maybe originating in nuclear complexity). Perhaps—and at this point I emphasize that this is only a shadowy suggestion—the problems within quantum chaos arise because chaos and quantum mechanics are treated on the same level. What if nonlinear dynamics and chaos were actually more fundamental than quantum mechanics?! This would mean that nonlinear corrections were being applied to themselves in a sense, with unanticipated results. This is one of the prime questions to be raised in this chapter.

Not only has the advent of research into quantum information theory [23] led to a renewed interest in investigating the foundations of quantum mechanics, but also it has led to more quantitative formulations of quantum mechanics as a statistical, probabilistic theory [24]. No one has seriously questioned the necessity for a statistical interpretation of quantum mechanics, but the difficulties—and paradoxes—arise when this interpretation is taken too literally and especially when statistical arguments are applied to individual events. Actually, during this past decade Tsallis and his co-workers have proposed an extended, "nonextensive" entropy for classical systems [25], which leads to a number of similarities with the statistics of quantum mechanics. In Section IV, this is considered in the context of an application to/extension of Bell's inequalities.

To summarize the problem succinctly: Although quantum mechanics has proven to be one of the most successful, most precise disciplines known, when one tries to understand its implications within the prevailing Copenhagen interpretation, one is quickly lead into a labyrinth of "mysteries, puzzles, and paradoxes"—counterintuitive imponderables. Most working scientists take these in stride, and, in fact, many take a perverse pride in expounding the peculiarities. Nonlinear corrections have been suggested a number of times; however, they remain exactly that—corrections to an essentially linear theory, which does not and cannot achieve chaotic behavior, which is fundamentally different. On the other hand, if one were to accept strongly nonlinear behavior underlying quantum mechanics, some of the so-called imponderables appear to have rational, deterministic explanations. Chaos theory itself is in its infancy, and its application to quantum mechanics is in a fetal stage. Thus, it is impossible to provide quantitative solutions at this point. Nor is that the intent of this chapter—the intent is to raise questions and perhaps provoke someone to delve more deeply into the myriad possible problems.

Because chaos theory is so new and so different from quantum mechanics, I felt it necessary to recall a minimum few of the relevant peculiarities of chaos theory in Section II. In the ensuing sections I show how parallel, nonlinear explanations can be formulated for several quantum mechanical imponderables. Then I summarize the questions raised and let the reader decide on whether or not they are worth investigating more deeply. As a start, Table I lists important comparisons among the four major disciplines of dynamics: classical dynamics, thermodynamics, quantum mechanics, and nonlinear dynamics. Here one can anticipate some of the similarities and relevancies of nonlinear dynamics to quantum mechanics.

## II. Some Relevant Peculiarities of Chaos

Although chaotic behavior was discovered by Poincaré at the end of the nineteenth century during his study of three-body planetary motion, its modern, quantitative rebirth had to wait until the work of Lorenz in the late 1950's [27]. And the surge of development leading to its becoming a mainstream science had to await the development of powerful modern computers and especially computer graphics to deal with and sort the plethora of numbers associated with numerical solutions of nonlinear maps and differential equations. This has come about only during the last several decades. Thus, the original formulators of quantum me-chanics did not have access to any well-developed theory of strongly nonlinear dynamics, nor did most of the later workers who suggested alternatives for or corrections to the Copenhagen interpretation.

**Table I.** Comparisons of Four Dynamical Systems

|  | **Classical Dynamics** | **Thermo-dynamics** | **Quantum Mechanics** | ***Nonlinear Dynamics*** |
|---|---|---|---|---|
| **Nonlinearity and Feedback** | normally no | possible | no by definition | yes |
| **Arrow of Time** | no | yes | ultimately yes [26] | definitely yes |
| **Statistical Behavior** | no | yes | yes | superficially yes |
| **Deterministic Behavior** | definitely yes | within statistics | no | in principal yes |
| **Isolated System** | preferably | not necessarily | no! à la Copenhagen | yes |
| **Self Organization** | no | yes far from equilibrium | no | yes |
| **Innately Preferred Modes (Quantization)** | no | no | yes | yes |
| **Parity Nonconservation** | no | no (NA) | yes in weak interactions | possible (e.g., odd iterators) |
| **Action at a Distance** | by means of fields | NA | yes | apparent |
| **Degree of Fundamentality** | none by convention (Correspondence Principle) | none by intention | extreme by convention | currently debatable |
| **How Logical Is It?** | extremely | logical in statistical limit | fundamentally counter-intuitive | after analysis, very |

It is worthy of note that many earlier writers on the fundamentals of quantum mechanics have figuratively danced closely around the notion that it contains similarities with nonlinear classical dynamics, without quite realizing the significance of such nonlinearities. For example, Feynman wrote [28]:

> Of course we must emphasize that classical physics is also indeterminate, in a sense. It is usually thought that this indeterminacy, that we cannot predict the future, is an important quantum-mechanical thing…It is true classically that if we knew the position and the velocity of every particle in the world, or in a box of gas, we could predict exactly what would happen. And therefore the classical world is deterministic. Suppose, however, that we have a finite accuracy, and do not know *exactly* just where one atom is, say to one part in a billion. Then as it goes along it hits another atom, and because we did not know the position better than to one part in a billion, we find an even larger error in the position after the collision. And that is amplified, of course, in the next collision, so that if we start with only a tiny error it rapidly magnifies to a great uncertainty…
>
> Speaking more precisely, given an arbitrary accuracy, no matter how precise, one can find a time long enough that we cannot make predictions valid for that long a time. Now the point is that this length of time is not very large…The time goes, in fact, only logarithmically with the error, and it turns out that in only a very, very tiny time we lose all our information. If the accuracy is taken to be one part in billions and billions and billions—no matter how many billions we wish, provided we do stop somewhere—then we can find a time less than the time it took to state the accuracy—after which we can no longer state what is going to happen!…For already in classical mechanics there was indeterminability from a practical point of view.

This was written in 1965, before the development of modern chaos theory, but it presages the "butterfly effect"—the supersensitivity of chaotic systems to initial conditions, with exponential divergence of orbits in phase space.

Perhaps even more to the point, in preparing for the justification of his "theory of the double solution," de Broglie comes close to recognizing the need for an underlying deterministic but "statistical in nature" theory such as chaos [29]:

> The usual interpretation of the formalism of wave mechanics is purely probabilistic, i.e., no attempt is made to look beyond the laws of probability which we have explained, and the idea of a hidden reality, on which the laws of probability are based, is rejected. This positivistic interpretation is based on the assertion that everything which is unobservable is non-existent and should have no place in theoretical physics. This assertion seems to me to be highly debatable. The fact that a particular entity is unobservable need not preclude its use in theoretical physics if it helps in our understanding of the subject and removes undesirable paradoxes.

He also accepted and made use of the quasi-thermodynamic random perturbations proposed by Bohm and Vigier [30] to introduce a random, probabilistic element into their hidden-variable formulation of quantum mechanics. (These came about supposedly because of the interaction of a particle with a hidden, subjacent "subquantum medium.")

Thus, some of the earlier formulators of quantum mechanics who were critics of the Copenhagen interpretation apparently recognized the need for nonlinear, perhaps chaotic behavior long before the formal theory of chaos started being developed. The field of nonlinear dynamics and chaos theory is vast and for the most part untapped, so here I introduce you to a few relevant concepts, chosen to illustrate their peculiar parallels to quantum behavior.

•**Innate modes and chaotic behavior in elementary, simple systems.** We are schooled to associate complex behavior with complex systems or, at the minimum, with outside or noisy influences on simple systems. However, complex, chaotic, seemingly random behavior can arise in the simplest of nonlinear systems. All that is required is some sort of feedback and at least second-order (quadratic) behavior.

The most studied simple system is the quadratic map,

$$x_{n+1} = x_n^2 + c, \tag{1}$$

introduced into population dynamics by Verhulst in its guise as the logistic map as early as 1844. (The two maps are related by a straightforward change of variable.) Simply stated, a population in generation $n + 1$ depends on the population in the previous generation $n$, and is related to it by the "birth rate," which can be taken as a control constant $A$,

$$x_{n+1} = Ax_n. \tag{2}$$

This leads to runaway, Malthusian growth, so it must be modified by taking into account the relationship of the current population to the maximum sustainable population, which is taken to have the value 1. Thus, the population $x$ is normalized to the interval [0,1], and the logistic equation results:

$$x_{n+1} = Ax_n(1 - x_n) \tag{3}$$

Here the growth rate is small when the population is small, but it is also small when the population nears the sustainable maximum, where the $(1-x_n)$ term tends toward 0.

This deceptively simple-looking equation exhibits unexpected, complex behavior. First of all, for the control parameter $A<1$, the population vanishes for all initial values. (We say that 0 is an *attractor* of the system.) For values of $A>1$, after sufficient iterations the population $x_\infty$ settles down onto a single value that depends on the specific value of $A$. Just above $A=3$ the final value of the population bifurcates, i.e., alternates between two stable end values—there are stable final modes, so to speak. As the value of *A* continues to increase, further bifurcations occur, with periods of 4, 8, 16, … (The periods are not just powers of 2 but incur all the natural numbers in an elaborate number-theoretical sequence called the Sarkovskii ordering [31].) Finally, at *A*>3.44948… chaotic behavior ensues, in which the map never settles down but continues hitting seemingly random values ad infinitum. The "dust" in Fig. 1 results not from poor graphics resolution nor from not carrying out sufficient iterations to fill in the gaps—it results from infinitely complex detail in the diagram, which is said to be "self-similar" (or more correctly, "self-affine"). In the mathematical limit of infinite iterations, the magnification—no matter how high, even infinite—of any portion of the diagram resembles the diagram as a whole. (There are regions in maps where every possible value is visited ("ergodic" behavior), whereby everything becomes completely filled in, but for this map such behavior occurs only at *A*=4.) Now, if this chaotic behavior were truly random, there would be little point in following through on chaos theory; however, there is a definite, albeit subtle order in chaos. One of the clearest manifestations of this can be seen in Fig. 1,where gaps of periodic order occur in the chaotic regions; one of the clearest is the large period-3 gap in the vicinity of *A*=3.82. There are an infinite number of these gaps, persisting all the way down to infinite magnification.

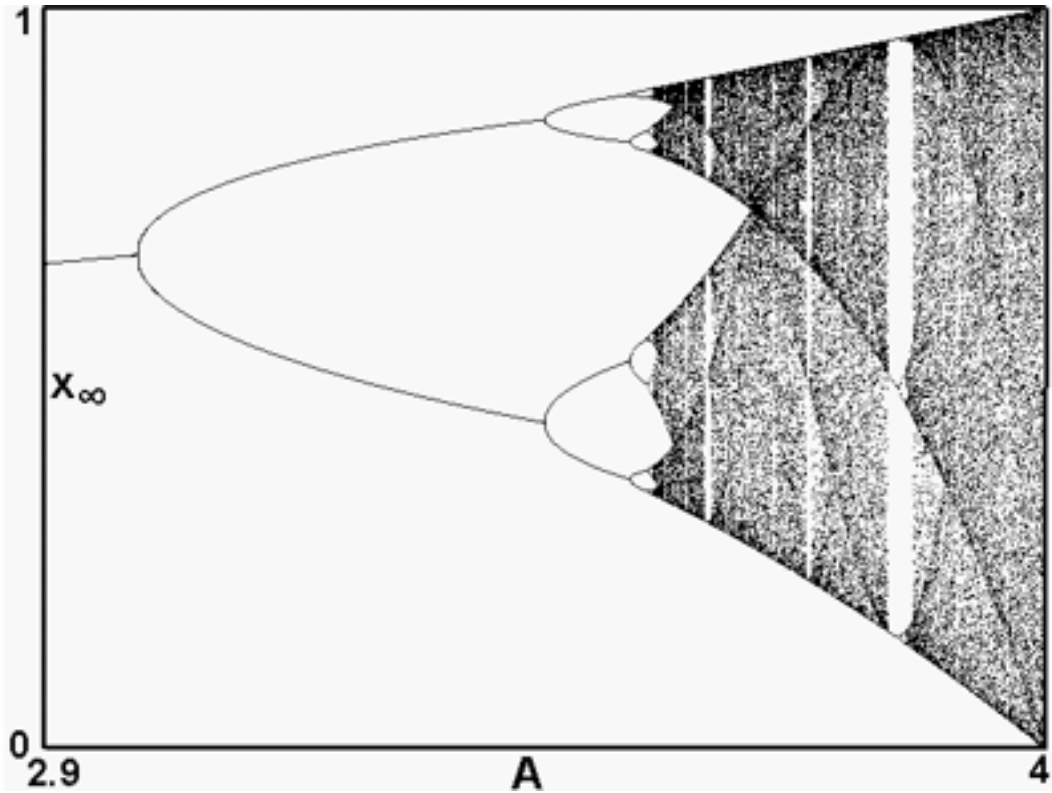

Fig. 1. Bifurcation diagram for the logistic map, where the final values of the population *x* are plotted against values of the control parameter *A* for the region of interest above 2.9. Bifurcations begin at *A*>3, with successive bifurcations leading to chaos at *A*>3.44948… For *A*>4, final values tend toward –∞ and the map breaks down.

•**Universality in chaos.** Again, inasmuch as nonlinear systems cannot be solved in closed form, if every nonlinear system were different, it would be a Herculean task to try to solve each system from scratch. There appear, however, to be universal traits common to many disparate nonlinear systems. For example, Feigenbaum discovered that entire classes of maps behave predictably in their bifurcation behavior [32]. For example, all unimodal maps (maps having a single smooth maximum and no "extraneous" inflection points—e.g., the Gaussian map with its inflection points is not unimodal and exhibits a very diifferent bifurcation diagram) share a common approximate symmetry: The ratio of the distance (in the value of *A*) between points where periods $2^{n+1}$ and $2^n$ are born to the distance between where periods $2^{n+2}$ and $2^{n+1}$ are born is a "universal" constant:

$$\delta = \lim_{n \to \infty} \left( \frac{A_{n+1} - A_n}{A_{n+2} - A_{n+1}} \right) = 4.66920161\ldots \tag{4}$$

As indicated by the equation, this relation holds more precisely as the number of bifurcations increases (which causes experimental problems, since most experiments can reach only a few low-order bifurcations). Thus, very different systems can be handled by similar mathematical analyses.

Universality is both a blessing and a curse. It is a blessing because it allows one to determine rather quickly whether or not a system is behaving chaotically, using techniques already developed for other systems. But it is also a limitation, because the similar behavior of entire classes of systems makes it difficult to sort out detailed mechanisms for a specific dynamical system.

•**Basins of attraction and preferred innate modes—extreme sensitivity to initial conditions.** Each nonlinear attractor can have its own basin of attraction—and this basin can range from the (perhaps point) attractor itself through straightforward Euclidean objects to extremely complex fractal objects. An example can make this clearer.

Fig. 2 shows the basins of attraction for a pendulum having three evenly-spaced attractors—for example, a long pendulum having an iron bob that is attracted by three magnets placed as the vertices of an equilateral triangle (each magnet placed at the center of its respective large-area basin). If the pendulum is started in the vicinity of a given magnet, then its final behavior is quite predictable. Its path may well seem erratic, but it will end up at rest above that same magnet. If, however, it is released near the equidistant line between two magnets, its behavior becomes much less predictable—sometimes it lands above one magnet, sometimes above the other, and sometimes above the distant magnet! A blow-up of a portion of the attractor shows increasing complexity. In the unpredictable regions, it can be shown [34] that every juncture between two basins must adjoin the third—and this increasing complexity continues as far down as we wish to magnify the diagrams. This is an illustration of extreme sensitivity of a chaotic system to initial conditions (termed the "butterfly effect" by Lorenz [35]), in which trajectories in phase space diverge exponentially. In some of these regions of extreme complexity it is impossible to determine the initial position of the bob with sufficient precision to be able to predict on which magnet it will finally wind up. This is also an example of a self-affine fractal (as was Fig. 1). Interestingly enough, the basins of attraction for Newton's method for extracting roots resembles that of a four-magnet pendulum and explains why Newton's method yields unpredictable results for certain initial guesses.

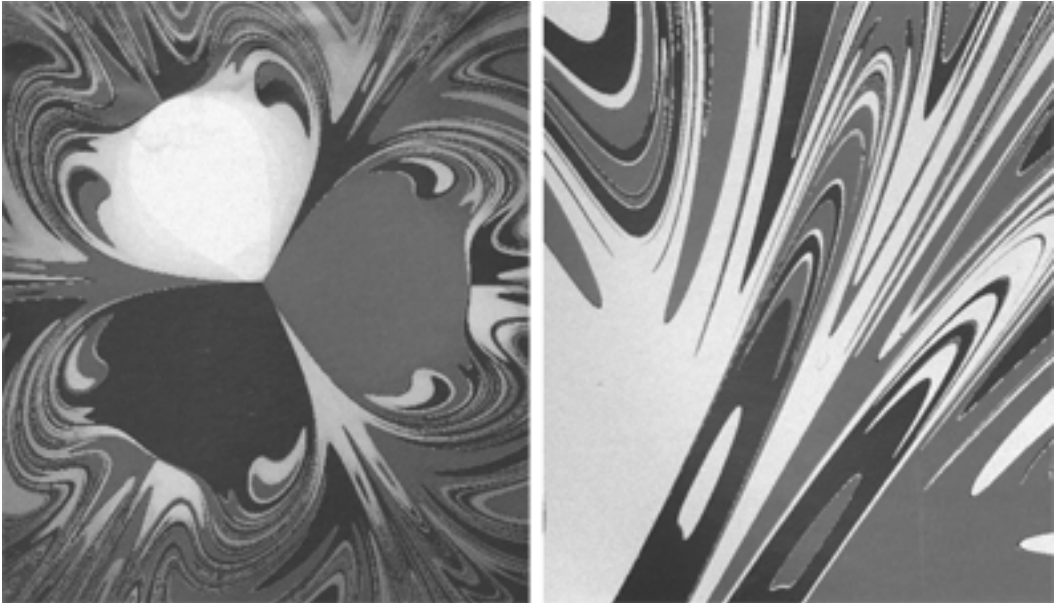

Fig. 2. Basins of attraction for a pendulum whose bob is attracted by three magnets placed at the vertices of an equilateral triangle (each vertex lies at the middle of one of the solid-shade areas). If the bob is released in the vicinity of one of the magnets, its final position will be above that magnet, but its final position quickly becomes less predictable when the bob is released from other locations. The right half of the figure is an enlargement of the area (upper right) between the large white and gray basins, and further enlargements show more and more interlocking areas. Between every two shades lies the third shade, ad infinitum. This is an example of extreme dependence on initial conditions (the "butterfly effect"), and in many regions it is impossible to know the initial position with sufficient experimental precision so as to be able to predict which attractor will win. (Adapted from Ref. [33].)

Attractors (and their basins of attraction) occur naturally for dissipative systems, where over time the systems settle down onto an attractor. The overall behavior of a system is linked to the properties of these attractors, which can be both stable (e.g., the point at the bottom of a pendulum's arc) and unstable (the point at the top of the arc, where the tiniest perturbation causes the bob to move away from that point)—or a combination of both (the simplest example being a saddle point, which is attractive from one direction and repulsive from another). In higher-dimensional phase spaces, generalized saddle attractors can form complicated topologies (homoclinic or heteroclinic tangles), whose interactions can give the appearance of "action at a distance," which could have some bearing on quantum mechanical problems.

•**Conservative systems—constants of motion and integrable Hamiltonians; dissolution of KAM tori.** The situation becomes less clear for Hamiltonian systems, but so-called "integrable" systems (which in Hamilton's formulation are dependent only on the action variables and not on the angle variables) exhibit constants of motion, i.e., can be naturally quantized [36]. These can rest on a knife-edge of stability, where small perturbations drive them into dissipative modes. An elaborate theory exists for describing this, involving so-called *KAM tori* and their progressive dissolution [37]. Weinberg [10] makes mention of these in his nonlinear discussion—and it could have ramifications for decoherence of quantum states.

•**Spontaneous symmetry breaking—parity nonconservation.** Much has been made of spontaneous symmetry breaking in particle physics, e.g., the acquiring of mass by vector bosons and parity nonconservation in weak interactions. But nonlinear systems have long been known to exhibit spontaneous symmetry breaking, sometimes at the apparent violation of the second law of thermodynamics. A practical example is the separation of powders in a nonlinear tumbler [38], in which a more or less homogeneous mixture "spontaneously" separates into component bands.

Odd-order maps also exhibit this phenomenon, as is illustrated in Fig. 3, which shows parts of bifurcation diagrams for the sine map,

$$x_{n+1} = A\sin(x_n). \tag{5}$$

Here the final values $x_\infty$ are shown for both positive and negative values of the initial parameter $x_n$, and at first glance these appear to be mirror images of each other. Upon further scrutiny, however, one finds subtle and not so subtle differences, especially in the chaotic regions. (It is not actually clear whether or not the chaotic regions are or are not mirror images, since there could be a fine interlacing of (mirror image?) points.) Odd-order maps such as this have not received detailed study or interpretation (this could well be a profitable field of research), but it seems clear that such maps do exhibit some spontaneous symmetry breaking, which could have bearing on, say, parity nonconservation in β decay and weak interactions.

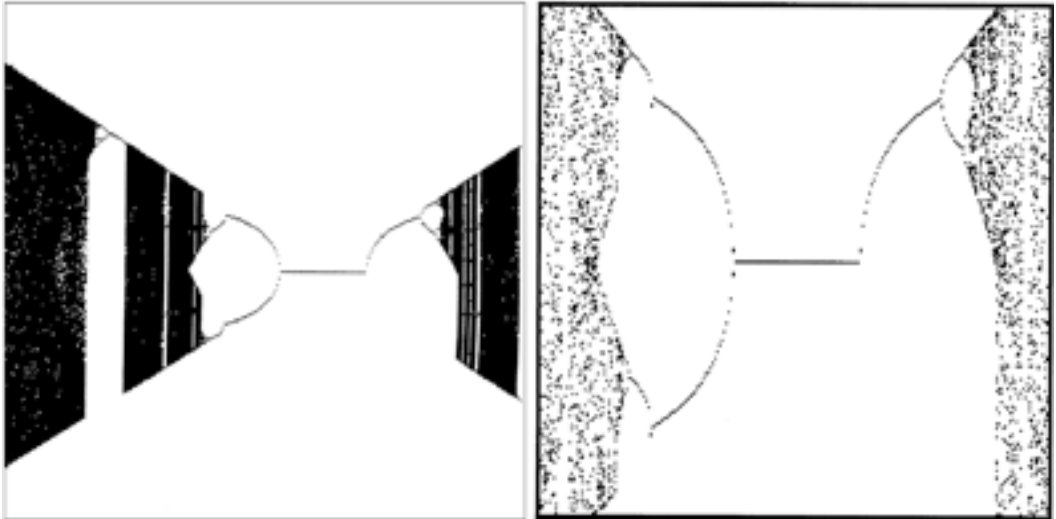

Fig. 3. Portions of the bifurcation diagram for the sine map, showing final values for both negative and positive initial values of $x_n$. At first glance these appear to be mirror images of each other, but closer inspection reveals differences. The diagram on the right is a massive blow-up of the central portion.

•**Order in chaos.** As mentioned above, if chaos were unpredictable, it would be of little interest. But it is highly predictable, and, among its more interesting phenomena is the appearance of order in chaos. In both Figs. 1 and 3 this can be seen. Once the bifurcation sequence reaches chaotic behavior, there still exist an infinite number of windows of order, in the form of periodic behavior. And it should be emphasized that this alternation continues down to infinite magnification. Thus, in extreme sensitive regions, chaos could produce not only exponential divergence of trajectories in phase space, but also an alternation between predictable (periodic) behavior and less predictable (chaotic) behavior. If one associates these with particle- and wave-like behavior, it could well have implications for duality in quantum mechanics.

In this section I have introduced only a selected few of the eccentric ideas from chaos, chosen for their possible relevance to parallels for quantum mechani-cal imponderables—the merest tip of the iceberg, so to speak. In the following sections I demonstrate several of these parallels in greater detail.

## III. Exponential Decay and Unimodal Maps

One of the more straightforward parallels in which a puzzling quantum mechanical phenomenon can be mocked up as a chaotic model is the exponential decay law. Over the years this decay law has been the subject of numerous investigations, both experimental and theoretical. Many experimental studies have focused on trying to detect deviations from exponential behavior either at very short or at very long times compared with a species' half-life [39]. Other, theoretical studies have tried to reconcile nonexponential predictions that come about for individual quantum states, such as a decaying nucleus, with the statistical exponential behavior of large ensembles [40,41]. To date no deviations from exponential behavior have been detected [39].

An empirical, statistical exponential decay law, having a time-independent half-life, appears to hold for any quantum system displaying first-order kinetics. These include unstable elementary particles, radioactive nuclei, atomic and molecular de-excitations, and even spontaneously decomposing molecules—any system in which the observed rate of activity or decay is directly proportional to the amount (or number) of disintegrating species at a given time. The governing first-order rate equation,

$$-\frac{dN}{dt} = \lambda N, \qquad (6)$$

has an integrated solution,

$$N = N_0 e^{-\lambda t}, \qquad (7)$$

where $N$ is the amount present at a given time $t$, $N_0$ is the amount present at $t = 0$, and the decay constant $\lambda$ is related to the half-life by

$$\lambda = \ln 2 / t_{1/2}. \qquad (8)$$

The noteworthy fact about first-order kinetics is that here and only here is the half-life time independent, so that it makes no difference where one chooses the starting $t = 0$; in all other cases the "half-life" is time dependent and thus not a particularly useful concept.

One common "practical" interpretation of the exponential decay law is by analogy with actuarial tables, such as those used by life insurance companies. It is impossible to predict the life expectancy of a single person; yet, given a large enough population, extremely precise statistical predictions can be made about overall life expectancies. By analogy, although it is impossible to predict when a particular radioactive nucleus will disintegrate, given a large enough ensemble, one can predict within statistical certainty that an exponential decay law will ensue, exhibiting the half-life characteristic of that particular species.

Upon further consideration, however, one finds that the analogy does not hold. Actuarial tables are based on complexity: Large numbers of differing people suffer from a myriad of different, unrelated causes of death, and these tend to average out, allowing the statistical predictions to have empirical validity. The minute that any correlations occur, as happens with natural disasters, epidemics, or war, the actuarial predictions break down and lose all validity. Contrasted with this, one of the fundamental concepts of quantum mechanics is that "identical particles" are indeed indistinguishable. All electrons are fundamentally the same, as are all radioactive nuclides of a particular $Z$, $N$, and isomeric configuration. Thus, one particular nucleus is absolutely indistinguishable from the next, which is indistinguishable from the next, which…—from this viewpoint they all should decay at the same time. Yet they decay at quite different times, producing the exponential decay curve.

It turns out that the extreme sensitivity of chaotic systems to initial conditions, *consistent with but not necessarily dependent on the Uncertainty Principle*, can provide a parallel analogy and might possibly lead to some insight into this problem. For iteration of unimodal maps can also lead to an exponential decay law from an ensemble of almost identical initial states, the key words here being "almost identical."

Because of the universality inherent in chaos, different unimodal maps should show the same basic behavior. In keeping with Ockham's Razor, we therefore choose the simplest, most studied map—the logistic map [Eq. (3)]. We choose an example with $A = 4$, not only chaotic but ergodic in behavior. And we use a "prisoner-escapee" procedure, as illustrated in Fig. 4.

Fig. 4, a plot of $x_{n+1}$ vs $x_n$, demonstrates the procedure of graphical iteration and escape. The parabola is simply a plot of Eq. (3) with $A = 4$, and the diagonal is the locus of points where $x_{n+1} = x_n$, enabling one to use the output from one iteration as the input of the next. From an arbitrary initial point $x_0$, one draws a vertical line up to where it intersects the parabola—this point of intersection is the next value $x_1$. Then a horizontal line is drawn to the diagonal, locating $x_1$ on the abscissa as the next input. Now, for our prisoner-escapee game, we need a

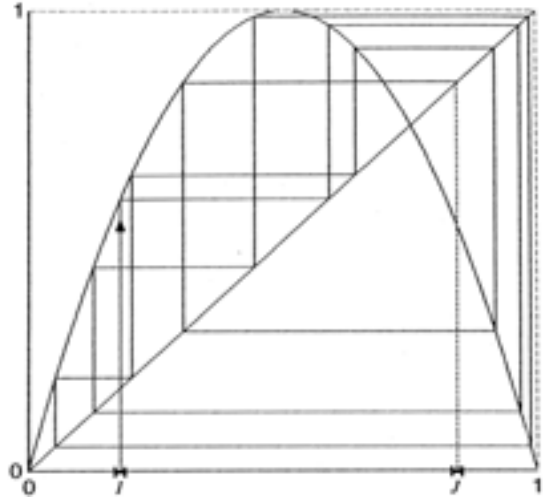

Fig. 4. Graphical iteration of the logistic map, demonstrating the "prisoner-escapee" procedure for generating an exponential decay law. Random points are selected from the initial interval *I*, which represents the initial decaying state. A vertical line up to the parabola locates the next point. A horizontal line over to the diagonal (where the ordinate equals the abscissa) takes this point into position for input into the next iteration. The procedure is continued until the trajectory "escapes" into the interval *J*, which corresponds to the final state. A record of the number of surviving states is kept, and these are plotted against the number of iterations to obtain an empirical decay curve.

tiny interval of almost identical states, so we choose points within the interval *I*. (These should be chosen randomly, but for this particular ergodic case with *A* = 4 equally spaced points should produce the same result.) Each initial point is iterated and a record kept of its trajectory until this trajectory "escapes," i.e., lands in a second interval *J*, corresponding to the final state. One then plots the number of initial points still remaining ("prisoners") after each iteration against the number of iterations to obtain an empirical decay curve.

A physical justification for this procedure is that a transition probability is in general at least a quadratic function. For example, the generic expression for the transition probability for emitting a γ ray from a nuclear de-excitation (for an electric multipole) is

$$B(El) = \langle \Psi_f | Op. | \Psi_i \rangle^2, \qquad (9)$$

which for a collective rotational electric quadrupole (*E*2) transition (say, in the ground-state rotational band of a deformed even-even nucleus—one of the more straightforward and better understood nuclear transitions) works out to be [42]

$$B(E2; K, J \to K, J') = \langle \Psi_{K,J} | M(E2) | \Psi_{K,J-2} \rangle^2. \tag{10}$$

Here $M(E2)$ is the sum of the components of the electric quadrupole moment with respect to the body-fixed symmetry axis of the nucleus, $J$ is the total nuclear angular momentum, and $K$ (= 0 for all states in the ground-state rotational band of an even-even nucleus) is its projection on the symmetry axis. The physical analogy of this with iterating the map is that an initial state $x_n$ located somewhere within the interval $I$ is taken by the iteration process until it can overlap with the final state $x_f$ located somewhere within the interval $J$; when this occurs the state can be said to have "escaped," i.e., successfully decayed to a final state within reach. The actual number of iterations can be likened to something such as the actual number of iterations of a real nuclear process—here, the number of oscillations of the electric quadrupole. (For $\alpha$ decay it could be likened to the number of attempts at barrier penetration by an $\alpha$ particle, or for $\beta$ decay it could be the number of oscillations of an intermediate vector boson before breaking up into an electron and neutrino—a less clear analogy.)

Fig. 5 shows the results from a computer experiment in which 10,000 points were selected from the tiny initial interval, [0.2, 0.2 + $10^{-11}$], and followed until their trajectories escaped into the final interval, [0.53, 0.54]. A well-defined empirical exponential decay curve results, having a half-life of about 107 iterations. (Naturally, the actual half-life obtained will depend sensitively on the placement and width of the two intervals.) Although it does not show up too clearly on this figure, there is an initial transient period, as might be expected from chaotic systems, before the curve settles down to exponential behavior.

Similar results should be expected from iterating other unimodal maps. An example of this from an experiment we performed with a slightly altered form of the sine map can be found in Ref. [43].

Time-delay series make powerful tools for analyzing chaotic behavior [44]. For a sequence of events separated by seemingly random time intervals, one makes a plot of interval $\Delta t_{m+n}$ vs interval $\Delta t_m$, where $n$ is the number of intervals delay between the successive pairs plotted. This procedure can show whether the data are truly chaotic or merely random, and it can bring out hidden order, sometimes helping to identify the mechanism that generated the chaotic behavior. The procedure works very well for mesoscopic systems, where the sampling time delay lies within a few orders of magnitude of the intrinsic physical process itself. For ultramicroscopic quantum systems, however, the sampling time would be many orders of magnitude greater than the intrinsic system iteration time, making *n* an enormous number—and making it unlikely that meaningful correlations could survive over such a large time interval. Nevertheless, such experiments (e.g., measuring delay times between fast cascading γ rays) should be very easy to perform, so they should be performed even though the chances of obtaining positive results are slim. It should also be noted that—if this scenario has any validity at all—the predicted transient times for observing nonexponential behavior should be so short as to be undetectable (cf. the searches outlined in Ref. [39]).

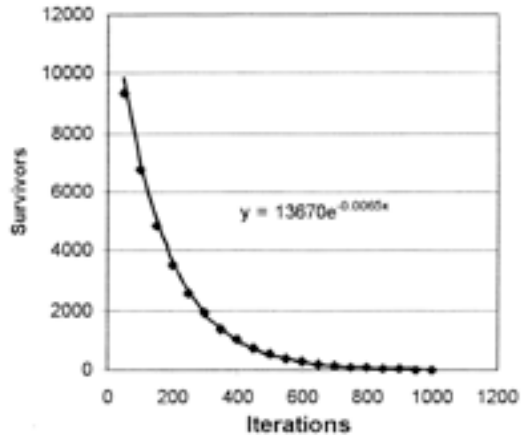

Fig. 5. Exponential decay curve produced by iterating the logistic map with $A = 4$. To represent the initial state, 10,000 evenly-spaced points were chosen from the interval, $[0.2, 0.2 + 10^{-11}]$, and the resulting trajectories were followed until they escaped into the interval, $[0.53, 0.54]$, representing the final state. This plot of the number of survivors vs the number of iterations produces an empirical exponential decay curve having a half-life of about 107 iterations.

This nonlinear, chaotic analogy to quantum mechanical decay, straightforward, fascinating, and perhaps even useful, must remain incomplete without rigorous mathematical proof—something most difficult to achieve in dealing with any sort of quantum mechanical phenomena. Nevertheless, it does illustrate that minute differences in initial quantum states—consistent with the Uncertainty Principle—can lead to significant differences in trajectories and consequently in behavior because of the extreme sensitivity of chaotic systems to initial conditions. Quantitatively, the parallel results in a statistical exponential decay law, consistent with the observed first-order kinetics of disintegrating quantum systems.

# IV. Bell's Theorem and Nonlinear Dynamics

Bell's inequality [14] lies at the heart of the revived interest in delving into the foundations of quantum mechanics. This came about because, for the first time, it and its subsequent variants provided experimentally meaningful possibilities to the abstract *Gedankenexperimenten* of the EPR paradox [2]. Bell's theorem and inequality deal with the statistics of measurements on separated EPR, i.e., entangled, pairs of particles. At first glance, classical mechanics dictates that the inequality should be obeyed, whereas quantum mechanics, using entangled states as a basis, allows violations. During the past several decades numerous experiments have been performed—and they all obtain results consistent with quantum mechanics and inconsistent with "classical" mechanics.

Interpretations have varied, but most of them invoke some idea that "local reality" must be absent from any so-called hidden variable extensions to quantum mechanics. In other words, Einstein's "spooky action at a distance" becomes a reality!

On the other hand, most recently (during the last several years), there have been attacks on the fundamental assumptions underlying Bell's inequality and its subsequent manifestations. This has become an extremely controversial topic, and debaters have lined up on both sides. Interestingly enough, the Italian "Mysteries, Puzzles and Paradoxes in Quantum Mechanics" workshops have become a bastion for Bell proponents—backed up by many of the experimentalists who have performed Bell-type experiments (who may be said to have a vested interest in the outcome)—whereas, the Swedish "Quantum Mechanics: Reconsideration of the Foundations" conferences have become an equally intense haven for the Bell antagonists. Before going into any of this, let us examine the simplest example of a Bell-type inequality.

## IV.1. A "Classical" Derivation

There have been numerous derivations and variants of Bell-type inequalities. Perhaps the simplest and most straightforward is that of Clauser, Horne, Shimony, and Holt, sometimes referred to as the CHSH inequality [18]. It was formulated specifically to make it easy to test experimentally, and it serves as a good illustration for the entire collection.

Consider the following experimental arrangement, as depicted in Fig. 6. A referee, usually referred to as Charlie in information theory, prepares pairs of particles that have binary properties, such as spin up vs spin down or perhaps horizontal vs vertical polarization. It does not matter how he prepares these pairs, so long as it is reproducible—for example, they could be two electrons or two photons. He sends one of each pair to Alice and the other to Bob, who are experi-mentalists separated by an effectively infinite distance—any measurements one makes can have no effect whatsoever on the other's measurements, nor can there be any instantaneous or short-term communication between them. However, they do have precisely synchronized clocks, such that they can perform pairs of measurements simultaneously.

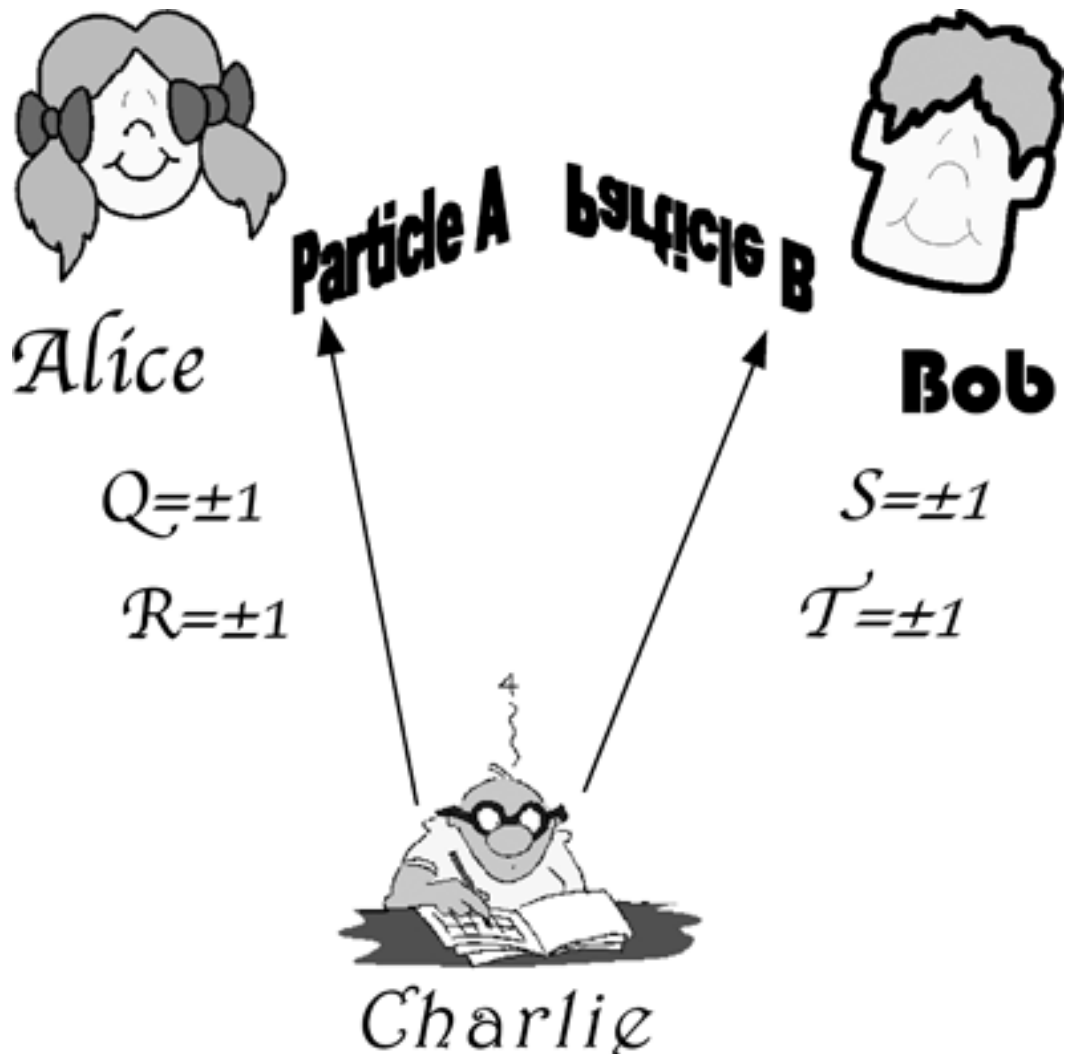

Fig. 6. Schematic for performing a Bell-type experiment on correlated pairs of particles. Charlie, the referee, prepares the pairs of particles (here implied to be a singlet or spins-opposite pair), then sends one particle of each pair to Alice and the other to Bob, who are experimentalists who cannot communicate with each other; however, their clocks are synchronized, allowing them to perform simultaneous measurements on their particles. Alice and Bob can each make one of two measurements, the outcome of which yields only +1 or −1, perhaps spin with respect to a vertical or an oblique axis, and each chooses randomly which measurement to make on a specific particle. After making many measurements (to achieve statistical validity), they get together to compare their results and to determine whether or not Bell's inequality (or the CHSH inequality) is obeyed by their results.

Alice is equipped to make measurements $Q$ or $R$ on each of her particles, each of which could result in an outcome of $+1$ or $-1$. For example, $Q$ could be a measurement of spin or polarization with respect to a vertical axis, while $R$ could be a similar measurement with respect to an oblique axis. Similarly, Bob can make measurements $S$ or $T$, which again can result in $+1$ or $-1$. And each randomly chooses which measurement to make on the particle from a given pair, perhaps deciding on which only after the particle is in flight, thus ensuring that there has been no communication with Charlie or the other experimentalist. After making many measurements in order to attain statistical significance, Alice and Bob finally get together to compare their results. The quantity of interest for them to compare is

$$QS + RS + RT - QT = (Q+R)S + (R-Q)T. \tag{11}$$

The single minus sign on each side of this equation is important. Because $Q$ and $R$ independently can take on the values of either $+1$ or $-1$, one or the other of the terms on the right side of the equation must equal 0. Either way, it is clear that

$$QS + RS + RT - QT = \pm 2. \tag{12}$$

After accumulating enough measurements to make use of statistics, Alice and Bob can speak in terms of probabilities. Letting $E(QS)$ be the mean value of the measurements obtained for the quantity $QS$, with corresponding expressions for mean values of the other products, we note that the mean values depend on efficiencies, errors in measurement, etc., such that they are always less than (or equal to in the ideal experiment) the theoretical values. We thus come up with the CHSH inequality, a more specific simplified version of Bell's inequality:

$$E(QS) + E(RS) + E(RT) - E(QT) \leq 2 \tag{13}$$

This "classical" derivation puts an upper limit on the statistical correlations on a *particular* combination of products obtained by presumably independent measurements on correlated pairs of particles made by experimentalists separated by an effectively infinite distance (i.e., with no possible communication or interactions between them). Let us now see what quantum mechanics has to say about a similar situation.

## IV.2. A Quantum Mechanical Derivation Involving Entangled States

Suppose that Charlie next prepares his pairs of particles in the quantum mechanical entangled Bell singlet state,

$$|\Psi\rangle = \frac{|01\rangle - |10\rangle}{\sqrt{2}}. \tag{14}$$

He sends the first qubit from each ket to Alice and the second to Bob, who proceed to make measurements as before, but now on the following specific observables:

$$Q = Z_1 \tag{15}$$

$$R = X_1 \tag{16}$$

$$S = \frac{-Z_2 - X_2}{\sqrt{2}} \tag{17}$$

$$T = \frac{Z_2 - X_2}{\sqrt{2}} \tag{18}$$

Here $X$ and $Z$ are the "bit flip" and "phase flip" quantum information matrices, corresponding respectively to the Pauli $\sigma_1$ and $\sigma_3$ spin matrices,

$$X = \begin{pmatrix} 0 & 1 \\ 1 & 0 \end{pmatrix} \tag{19}$$

and

$$Z = \begin{pmatrix} 1 & 0 \\ 0 & -1 \end{pmatrix}. \tag{20}$$

These may seem like arbitrary combinations, but remember that finding a single specific counterexample to a theorem is sufficient to disprove it. It is straightforward to show that the expectation values of the pairs $QS$, $RS$, and $RT$ are all

+1/√2, while that of $QT$ is −1/√2. As a result, we obtain the quantum mechanical version analogous to the CHSH inequality:

$$\langle QS \rangle + \langle RS \rangle + \langle RT \rangle - \langle QT \rangle = 2\sqrt{2}. \tag{21}$$

Quantum mechanics, within the framework of entangled states, predicts a larger number for the statistical correlation of the expectation values than was predicted by the so-called classical correlation on mean values. In other words, Bell's inequality places an *upper bound* on the predicted measurements, then quantum mechanics predicts a possible violation of this upper limit. It should also be noted that Eq. (21) predicts the *maximum* violation of Bell's inequality, which occurs only for certain angles (between the $Q$ and $R$ or $S$ and $T$ measurement axes) in specific experiments—so the experimental conditions must be chosen carefully beforehand.

A word about the current physical interpretation of these Bell-type experiments using Bell singlet entangled state. A Bell singlet state, as represented by Eq. (14), is a specific linear combination example of the general singlet state, in which, say, two spin-1/2 particles have their spins pointing in opposite directions. These could be two leptons emitted from a nuclear process. The important point is that *the specific directions in which the spins are directed is unknown*—all that is known is that, whichever direction the spin of the first particle points, the spin of the second particle must be pointing in the opposite direction. This is the physical state of affairs when particles from a pair are sent to Alice and to Bob. "Entanglement" means that the states of the two particles must be considered as a whole—the overall wavefunction cannot be factored into its respective parts. Physically, this means that the particles retain some sort of connected inter-relationship. According to the Copenhagen interpretation, one cannot speak about the individual spin directions—the particles exist in some sort of limbo, where the directions of their individual spins do not exist except for the fact that they must be antiparallel. However, when, say, Alice makes a measurement, she forces a "collapse of the wavefunction" such that the direction of her particle suddenly becomes well-defined in a specific direction. And this somehow forces Bob's corresponding particle to take on (a better term here than "line up" in) the opposite direction *instantaneously*! If necessary, faster than the speed of light! This sort of behavior is what Einstein termed "spooky action at a distance." But, according to Bell's theorem with all the variants of Bell's inequality, quantum mechanics (with its implied instantaneous action at a distance) can yield a different, possibly higher numerical correlation on statistical correlation from measurements made on such correlated, entangled particle pairs.

During the last several decades numerous Bell-type experiments have been performed, beginning with the early experiments of Clauser and Freedman

[45], continuing the more highly publicized experiments of Aspect et al. [46], and culminating with experiments having separations as great as 10 km between the detectors (utilizing fiber optics for signal transmission) [47]. A comprehensive coverage of these can be found in Ref. [17]. And essentially all of these experiments support the quantum mechanical predictions!

## IV.3. Interpretations, Other Variants, and Refutations

Interpretations concerning the violation of Bell's inequality and the consequent dominance of quantum mechanics over classical mechanics vary, but almost inevitably at some point they reach conclusions such as "local reality is ruled out in quantum mechanical systems." And at times they go so far as to rule out effectively any hidden variables constructs. Einstein's "spooky action at a distance" indeed becomes a reality!

Of the many variants of Bell's inequality, one of the more fascinating is the so-called GHZ formulation, which eliminates inequalities altogether [48]. It does this by extending the number of correlated particles to three or four. In the two-particle versions of Bell's inequality, such as the CHSH inequality discussed in the previous section, the statistical interpretation comes about because of the randomness in choosing the measurement directions. In those events where both Alice and Bob happen to choose the same axis in which to measure a particular pair (termed a "perfect correlation"), then the outcome of, say, Bob's measurement is certain once the result of Alice's measurement is known—this in keeping with the singlet nature of the Bell state. GZH show that for four entangled spin-1/2 particles (later adapted to three so-called spinless particles, where direction or momentum is the operative property to be measured) it is impossible to find experimental conditions (i.e., proper phase shifts introduced into certain paths) that can produce non-contradictory "perfect correlations." They thus concluded that resorting to statistical inequalities was unnecessary.

Their *Gedankenexperiment* involved three photons emitted at angles of 0, $2\pi/3$, and $4\pi/3$ or through a second set of apertures offset by an angle $\alpha$ (i.e., at $\alpha$, $2\pi/3 + \alpha$, and $4\pi/3 + \alpha$); the second set of paths also included phase shifters, e.g., glass plates. Thus, after passing through the apertures the overall state of the particles would be

$$|\Psi\rangle = (1/\sqrt{2})[|a\rangle_1|b\rangle_2|c\rangle_3 + |a'\rangle_1|b'\rangle_2|c'\rangle_3], \tag{22}$$

where $|a\rangle_1$ denotes particle 1 in beam *a*, etc., and the primes denote the offset beams. Beyond the apertures the beams are reflected so as to overlap at 50-50 beam splitters, and the outgoing beams are monitored by detectors *d* and *d'*, etc.

The phase shifters included in the offset paths introduce a variable phase shift $\phi$. Thus, the evolution of the kets is given by

$$|a\rangle_1 \to (1/\sqrt{2})[|d\rangle_1 + i|d'\rangle_1] \tag{23}$$

and

$$|a'\rangle_1 \to (1/\sqrt{2})e^{i\phi}[|d'\rangle_1 + i|d\rangle_1], \tag{24}$$

where $|d\rangle_1$ denotes particle 1 directed toward detector $d$, $i$ results from reflection at the beam splitter, and $\phi_1$ is the phase shift introduced into beam $a'$ (plus equations mutatis mutandis for the other beams). When these are combined with the state of Eq. (22), a state having eight terms results, from which the probabilities of detection of the three particles by the two set of three detectors can be obtained:

$$P_d^\Psi(\phi_1,\phi_2,\phi_3) = \tfrac{1}{8}[1 + \sin(\phi_1 + \phi_2 + \phi_3] \tag{25}$$

and

$$P_{d'}^\Psi(\phi_1,\phi_2,\phi_3) = \tfrac{1}{8}[1 - \sin(\phi_1 + \phi_2 + \phi_3)], \tag{26}$$

where the plus sign applies for an even number of phase shifts, the minus sign, for an odd number. If the expectation value is denoted as +1 for a particle entering an unprimed detector and −1 for entering a primed detector, then the expectation value of the product of the three outcomes is

$$\boldsymbol{E}^\Psi(\phi_1,\phi_2,\phi_3) = \sin(\phi_1 + \phi_2 + \phi_3). \tag{27}$$

"Perfect correlations" are finally obtained for the sums of angles,

$$\phi_1 + \phi_2 + \phi_3 = \pi/2 \quad (\boldsymbol{E}^\Psi = +1) \tag{28}$$

and

$$\phi_1 + \phi_2 + \phi_3 = 3\pi/2 \quad (\boldsymbol{E}^\Psi = -1). \tag{29}$$

GHZ then proceed to show that such a set cannot be found without contradiction, so by using the extra degree of freedom from three-particle correlations, they obtained a version of Bell's theorem "free from inequalities." This is the ideal *Gedankenexperiment*, however, and for projected real experiments they are forced to fall back on inequalities (which may be violated) for various combinations of phase shifters. In addition, a real test of their premises would require very specific (and exotic) cascades such as $J = 0 \to J = 1 \to J = 1 \to J = 0$, from an atom, with (hopefully) sufficient photons emitted toward the axes of a planar equilateral triangle as to make an experiment feasible—or perhaps a suitable triplet obtained from cascading down conversions.

A "simplified" variant of Bell-like theorems, much in the spirit of GHZ, has been promulgated by Mermin [49], and this type of variant has been discussed extensively in a philosophy/physics book by Albert [50]. The Mermin variant has also been the starting point for currently active attacks on Bell's theorems in general.

Hess and Philipp, for example, have attacked Bell's theorem in a series of papers [51], in which time-dependent "hidden variables" are introduced. Inasmuch as the electrodynamics of moving bodies cannot be described by time-independent theories, they conclude that Bell-type theorems simply are not suitable and cannot explain the dynamics of EPR-type experiments. Their conclusions have been refuted by Mermin himself [52] and by others [53], who have concluded that the Hess-Philipp propositions themselves are non-local in nature. The refutations have been refuted in turn [54], and this has all made for lively dialogs, most of which have been posted on arXiv and show few signs of abating.

## IV.4. Bell's Theorem Meets Nonlinear Dynamics

Returning to the CHSH inequality of §IV.1, we find that it is not the quantum mechanical predictions that are at fault. Instead, it is the so-called "classical" derivation that is suspect. Although Charlie supposedly prepared correlated pairs of particles, the derivation was actually carried through with a tacit suppression of correlations. Any correlations in the statistics would easily raise the upper bound—and nonlinear dynamics, even when not operating in a chaotic regime, would be expected to introduce correlations into the ensemble of events. Chaos itself often yields exponential rather than Gaussian distributions—for example, in the widespread $1/f$ systems [55] and also in self-organizing systems [56]. Such distributions fall off considerably more slowly from the mean than do Gaussian distributions. One might expect the "classical" version of Bell's inequality to be altered considerably if such correlated statistics were included in its derivation, because the slower fall-off means that "longer range" correlations (in this case referring to events lying far out in the tail, i.e., events unusual when compared with the norm) would have a greater effect on the statistics.

If I may be forgiven for introducing it, a trivial but rather flamboyant example illustrates this point. When I was a graduate student at Berkeley in the 1960's, a legal-statistical issue briefly caught the headlines. An old lady was mugged in Los Angeles, but witnesses were able to furnish a description of her attacker: a bearded, red-headed hippie man who fled into a yellow convertible driven by a long-stringy-haired blonde woman. The police soon found their suspect, who was convicted on the basis of statistics. I forget the exact numbers, but they went something like this: perhaps one person in a thousand in L.A. at that time was a red-bearded hippie, one in, say, one hundred was a long-haired blonde woman, and one in two thousand drove a yellow convertible automobile. Thus, that description statistically should fit only (1/1000)(1/100)(1/2000) or one out of two hundred million persons—not much doubt that they had the culprit! Not much doubt, that is, until someone pointed out that these were not independent statistics. A red-bearded hippie would undoubtedly choose to hang out with someone strange, such as a long-stringy-haired woman, and they would quite likely choose an unusual car, perhaps a yellow convertible. As a result, the statistics became correlated, lowering the odds considerably, so the conviction was thrown out. In scientific terms, "the upper bound on the statistical correlation was raised."

This type of reasoning—the introduction of correlation statistics into classical systems—has been codified by Tsallis and his co-workers [57, 58] in a new extension to entropy, which fits into a new extension to thermodynamics that they term "nonextensive" (meaning nonadditive) thermodynamics. Since its introduction some fifteen years ago, it has seen applications in many fields [59], and there have been a number of international conferences and workshops on the subject [60].

The definition of this generalized entropy is

$$S_q \equiv \frac{1 - \sum_{i=1}^{W} p_i^q}{q-1}; \quad (q \in \Re). \tag{30}$$

Here the phase space $\Re$ has been divided into $W$ cells of equal measure, and the probability of being in cell $i$ is $p_i$. For the exponent (termed the "entropic index") $q = 1$, this reduces to the usual entropy,

$$S_1 = -\sum_{i=1}^{W} p_i \ln p_i. \tag{31}$$

As $q$ differs from 1, the deviation from standard distributions becomes greater as well, with "long-range" correlations becoming more and more apparent.

Systems having $q > 1$ are termed "superadditive" (superextensive), while those with $q < 1$ are "subadditive" (subextensive), and the important point is that there is a cross-term (interference term) when such entropies are added:

$$\frac{S_q(A+B)}{k} = \frac{S_q(A)}{k} + \frac{S_q(B)}{k} + (1-q)\left(\frac{S_q(A)}{k}\right)\left(\frac{S_q(B)}{k}\right) \qquad (32)$$

Despite some attempt to justify nonextensive entropy theoretically [61], it remains basically an empirical—albeit increasingly useful—concept. It has been applied to everything from the distributions of wind velocities in tornadoes to the energy distribution in cosmic rays and from heavy nuclear-nuclear collisions. And it points out quantitatively that classical and macroscopic systems can exhibit apparent long-range correlations, especially in self-organizing systems. The causes are subtle and unclear, but it could well have to do with interactions between global attractors in nonlinear or chaotic systems.

It seems fairly clear that the use of such nonextensive entropies in the context of Bell's inequality could easily raise the upper bounds on classical correlations, such as those in the CHSH inequality above. This could well bring the classical predictions in line with the quantum mechanical entangled predictions, which would mean that the tests of Bell's inequalities do not rule out classical—but linear!—dynamics. The question of the existence of local reality then becomes moot—not because instantaneous long-range interactions have been proven to exist in classical dynamics, but because nonlinear systems have subtle ways of developing long-range correlations from local interactions.

There have also been somewhat parallel lines of thought by those working in information theory. Arguing in analogy with thermodynamics, entanglement has been likened to energy as a physical resource [62]. The science again is only in its first stages, but it is suggested that entanglement flow between systems can be governed by a law somewhat analogous to the second law of thermodynamics [63], and the degree of entanglement could be characterized by an information theoretical equivalent of entropy (Shannon entropy perhaps, but not necessarily leading to von Neumann entropy?).

Finally, an argument has been made by Tommasini [64] involving a statistical interpretation of quantum field theory, which shows that the standard model prevents the existence of entangled states having a definite number of particles. Within such an interpretation, Bell's theorem again does not forcibly demonstrate nonlocality.

In summary, Bell's theorem with its associated inequality(ies), has held sway for several decades, and Bell-type experiments supposedly have demonstrated the ascendancy of quantum mechanics and the demise of classical mechanics in interpreting the results. However, it now appears that Bell's theorem rests on various subtle assumptions, such as time-independent correlations and non-correlated statistics, which lend some doubt to its validity in interpreting EPR phenomena. Classical, nonlinear systems are far more subtle than once thought.

# V. Other Possible Parallels

There are other possible parallels in nonlinear dynamics to quantum mechanical paradoxes and imponderables, not so well developed. I list several of these here, in order of increasing speculativeness:

•**Attractors, basins of attraction, and implications for quantization.** This has been touched on already in the Introduction, but it may seem more plausible at this point. Nonlinear systems have their own preferred modes of oscillation, *independent of external influence*. This situation can be formalized by the existence of attractors and basins of attraction, which means that nonlinear systems can find themselves quantized without having to invoke, say, constructive and destructive wave interference. In other words, many deterministic but nonlinear classical systems innately satisfy eigenvalue equations.

•**Spontaneous symmetry breaking—parity nonconservation.** Nonlinear systems can spontaneously break both temporal and spatial symmetry. Good practical applications of this, as illustrated in §II, are the separations of powders in nonlinear tumblers [38]. The two regimes in which parity nonconservation occur in nature are weak interaction and in chemical (biologically active) compounds involving chiral atoms, usually chiral carbon atoms—and these two could well be related [65]. Might not quantum systems which violate parity, i.e., which have intrinsic handedness, have some sort of nonlinear basis? These could be mocked up, for example, by odd iterators (cf. the discussion of the sine map in §II).

•**Decoherence and the destruction of KAM tori.** One of the more extreme ideas resulting from the Copenhagen interpretation of quantum mechanics is the idea of decoherence, which can lead to implausible extremes, such as the Schrödinger's cat paradox [66]. The nonlinear parallel is that an observer could perturb, ever so slightly, a Hamiltonian system, resulting in its "decoherence," perhaps through the orderly breakdown of the KAM tori (cf. §II) into a dissipative system. This parallel is intuitively appealing, inasmuch as it removes the observer from being an essential part of a (closed) system. Again, Weinberg [10] touches upon this in his consideration of nonlinear effects in quantum mechanics.

•**Diffraction—order in chaos.** In many chaotic systems there exist regions where order and chaos are intimately mixed, right down to infinite magnification, as was demonstrated with the bifurcation diagrams and the basins of attraction for the three-magnet pendulum. As emphasized by Feynman [67], the double-slit experiment contains the essence of all that is weird about quantum mechanics. A possible nonlinear attack on this problem would be the following: Because chaos cannot ensue unless one finds at least quadratic governing equa-tions, one might take advantage of the order in chaos that occurs in many chaotic systems. A particle passing through a single slit would have its motion governed strictly by a first-order differential equation. However, when two slits are open, the passage of the particle through each slit could be governed by analogous first-order equations—and these could be combined to produce a single second-order differential equation, with the possibility of chaotic behavior. If so, the motion of the particle would be quite sensitive to position, and this could result (in an order in chaos realm) in an alternation between predicted order and chaotic order—in other words, something approaching a diffraction pattern might well be produced. The instant one of the slits was closed or a measurement made as to which slit through which the particle passed, however, then the behavior would revert to a single first-order equation, i.e., classical particle behavior. This is a current field of research [68], but finding the proper equations for producing quantitative results turns out to be a subtle, difficult task, made not easier by the masking universality of chaotic systems.

## VI. Conclusions

This chapter was intentionally meant to be somewhat provocative. In it I have raised the question of the possible existence of parallel nonlinear (possibly chaotic) explanations for some familiar quantum mechanical imponderables. Two of these, the exponential decay law and Bell's theorem have been discussed rather extensively—the other ones, much more speculatively. It should be emphasized that this has been an empirical—even experimentally minded—approach. Such reasoning does not provide a solid link between nonlinear dynamics and quantum mechanics, much less a proof that quantum mechanics indeed does have nonlinear, chaotic underpinnings. Indeed, it can be argued that, just as it is not possible to derive but rather only to postulate the principles of quantum mechanics, such a quantitative basis probably could never be proven. Quantum mechanics, for all its precision, success, and glory, really is fundamentally an empirical, statistic science. Nevertheless, these illustrative parallels should provide cause for serious consideration of the raised questions—and if sufficient quantitative parallels can be formulated, then this should motivate a more formal, mathematical search for the linkages. Mielnik's quotation [12] in the Introduction should receive serious consideration—we have been immersed for so long in the Copenhagen interpretation that we reflexively fight any questioning of it.

An important point to keep in mind is that supplying nonlinear underpinnings for quantum mechanics is not a hidden variables theory. Hidden variable extensions were invented so as to bridge the gap between the statistical nature of quantum mechanics and a sought-after more deterministic, more fundamental theory. Nonlinear dynamics in its chaotic realm does exactly this without having to invent extra, hidden variables. Fundamentally it is deterministic—there is a one-to-one correspondence between cause and effect—so this should make the followers of Einstein happy. On the other hand, because of its extreme sensitivity to initial conditions and the exponential propagation of errors, it is fundamentally impossible—except in the limit of infinite precision, which, of course, we can never attain—to predict with certainty the exact path of a given trajectory in phase space or of a specific measurement. Nevertheless, even if we are not tracking the trajectory we set out to track, we can rest assured that we are "shadowing" a nearby trajectory, which makes a statistical interpretation of quantum mechanics meaningful. As a result, it supplies the practical statistical interpretation that should also make the followers of Bohr content. It appears that, when all is said and done, both Einstein and Bohr could have been correct! Historically, they simply lacked the chaotic connection.

What might be origins of the nonlinearities? Once I thought that these might originate with relativistic effects—after all, with any reasonable potential, the Klein-Gordon (relativistic Schrödinger equation) becomes nonlinear [68]. This might be a possible source of nonlinearities, but upon reconsideration, I find it more likely that the measurement process itself is more likely the culprit. The "old-fashioned" interpretation of the Heisenberg Uncertainty Principle (as opposed to the now in vogue Copenhagen interpretation) might have some validity: The quantum mechanical description of a microscopic system could, after all, be deterministic, but any possible measurements that we can make on it forcibly drive it into a statistical regime.

Permit me one final, possibly naïve metaphor. An atom might be likened to an astronomical solar system. We suspect that our calculations on its future should fall into the chaotic regime, where infinitesimal initial differences could drive the entire system into seeming unpredictability. Perhaps we cannot predict the position of the comet a million years hence to better than the diameter of that solar system. That does not mean that, come one million years from now, the comet will be in actual limbo as to position. It will have a most definite position—we simply cannot calculate what it will be to much accuracy. An electron within the atom could well be in a parallel but more formidable situation. Not only is it impossible to calculate its exact position, but also, when the time comes, we cannot measure its position without massively perturbing it. Many of the imponderables, mysteries, puzzles, and paradoxes of quantum mechanics undoubtedly arise from this confusion of the possibility of calculation with that of real position—and many result from combining the statistics of a quantum mechanical ensemble with the behavior of a purported single particle or entangled pair of particles.

What might be the effects on quantum mechanics in practice if these suggestions have any validity?  Probably relatively little, especially at first.  There are multitudinous mathematical descriptions of almost any physical problem.  Remember, the old Ptolemaic epicycle description of the motion of the planets worked—predictions with good precision could be obtained.  It was an excellent example of a cumbersome theory, however—one that defied Ockham's razor, and we abandoned it in lieu of the far more believable Copernican model.  Now, calculations involving nonlinear dynamics and chaos are every bit as difficult as those for wave mechanics in its present day form.  So there should be little if any calculational trade-off.  And, as mentioned in the Introduction, few practicing quantum physicists worry about such things on a day to day basis, anyway.  However, there conceivably could be important ramifications for quantum computing, which is firmly based on the premise of linear superpositions.  If these were to fall by the wayside, quantum computing (along with quantum teleportation, etc.) could well be futher off than we think.  Or, on the other hand, there just might exist certain nonlinear parallels to linear superposition that could make massively parallel quantum computers feasible.

No firm conclusions—but many significant questions.  I am not naïve enough to think I have convinced or converted the reader to these propositions, but I do hope that she/he will ponder them further.  Quantum mechanics at the beginning of the 21$^{st}$ century may be an established science, but it is most certainly not a static, solved science.

# References:

[1] M.J. Feigenbaum, Preface to H.-O. Peitgen, H. Jürgens, and D. Saupe, *Chaos and Fractals: New Frontiers of Science* (Berlin: Springer-Verlag, 1992).
[2] A. Einstein, B. Podolsky, and N. Rosen, *Phys. Rev.* **47**, 777 (1935).
[3] N. Bohr, *Phys. Rev.* **48**, 696 (1935).
[4] Covered extensively in J.A. Wheeler and W.H. Zurek, Ed., *Quantum Measurement* (Princeton, NJ: Princeton Univ. Press, 1983).
[5] R. Bonifacio, B. G. Englert, and D. Vitali, Ed., *Proceedings of the 3$^{rd}$ Workshop on Mysteries, Puzzles and Paradoxes in Quantum Mechanics, Z. Naturforsch.* **a56** (2001); R. Bonifacio and D. Vitali, Ed., *Proceedings of the 4$^{th}$ Workshop on Mysteries, Puzzles and Paradoxes in Quantum Mechanics, J. Opt. B: Quantum Semiclass. Opt.* **4** (4), S253 (2003).

[6]  A. Khrennikov, Ed., *Quantum Theory: Reconsideration of Foundations, Conference Proceedings* (Växjö, Sweden: Växjö Univ. Press, 2001); A. Khrennikov, Ed. *Quantum Theory: Reconsideration of Foundations 2, Conference Proceedings* (Växjö, Sweden: Växjö Univ. Press, 2003).

[7]  L. de Broglie, *The Current Interpretation of Wave Mechanics: A Critical Study* (Amsterdam: Elsevier Publ., 1964); L. de Broglie, *Non-Linear Wave Mechanics: A Causal Interpretation* (Amsterdam: Elsevier Publ., 1960); and refs. therein.

[8]  D. Bohm, *Phys. Rev.* **85**, 166 (1952); also discussed in [4].

[9]  H. Everett III, *Rev. Mod. Phys.* **29**, 454 (1957).

[10] S. Weinberg, *Ann. Phys. (NY)* **194**, 336 (1989).

[11] N. Gisin, *Helv. Phys. Acta* **62**, 363 (1989); N. Gisin, *Phys. Lett.* **A143**, 1 (1990).

[12] B. Mielnik, *Phys. Lett.* **A289**, 1 (2001).

[13] M. Czachor and H.-D. Doebner, *Phys. Lett.* **A301**, 139 (2002).

[14] J.S. Bell, *Rev. Mod. Phys.* **38**, 477 (1966); J.S. Bell, *Physics* **1**, 195 (1964); these are reprinted in J.S. Bell, *Speakable and Unspeakable in Quantum Mechanics* (Cambridge: Cambridge Univ. Press, 1993).

[15] J. von Neumann, *Mathematical Foundations of Quantum Mechanics* (Princeton, NJ: Princeton Univ. Press, 1955), Chapters V and VI [translation of the original German edition of 1932].

[16] D. Bohm, *Quantum Theory* (Englewood Cliffs, NJ: Prentice-Hall, 1951), Chapter 22.

[17] R.A. Bertlmann and A. Zeilinger, Ed., *Quantum [Un]Speakables* (Berlin: Springer-Verlag, 2003); the history and interpretations are covered extensively in this book; Chapters 6 (by J.F. Clauser) and 9 (by A. Aspect) are particularly informative.

[18] J. Clauser, M.A. Horne, A. Shimony, and R. Holt, *Phys. Rev. Lett.* **23**, 880 (1969).

[19] P.A.M. Dirac, *The Principles of Quantum Mechanics*, 1st Edition (Oxford: Oxford Univ. Press, 1930); 4th Edition (Oxford: Oxford Univ. Press, 1958).

[20] H.-J. Stöckmann, *Quantum Chaos: An Introduction* (Cambridge: Cambridge Univ. Press, (1999).

[21] J. Ford, Chap. 12 in P. Davies, Ed., *The New Physics* (Cambridge: Cambridge Univ. Press, 1989).

[22] V. Zelevinsky, *Rev. Mex. Phys.* **48,S2** 18 (2002).

[23] M.A. Nielsen and I.L. Chuang, *Quantum Computation and Quantum Information* (Cambridge: Cambridge Univ. Press, 2000).

[24] L. Hardy, quant-ph/0101012 (2001).

[25] C. Tsallis, *J. Stat. Phys.* **52**, 479 (1988).

[26] M. Gell-Mann and J. Hartle, in J.J. Halliwell, J. Pérez-Mendez, and W.H. Zurek, Ed., *The Physical Origin of Time Asymmetry* (Cambridge: Cambridge Univ. Press, 1994).

[27] E. Lorenz, *J. Atm. Sci.* **20**, 130 (1963).

[28] R.P. Feynman, R.B. Leighton, and M. Sands, *The Feynman Lectures on Physics, Vol III: Quantum Mechanics* (Reading, MA: Addison-Wesley, 1965), p. 2-9.

[29] L. de Broglie, *The Current Interpretation of Wave Mechanics: A Critical Study*, p. 20.

[30] D. Bohm and J.P. Vigier, *Phys. Rev.* **96**, 208 (1954).

[31] R.C. Hilborn, *Chaos and Nonlinear Dynamics*, 2$^{nd}$ Edition (Oxford: Oxford Univ. Press, 2000), Chap. 5.

[32] M. Feigenbaum, *J. Stat. Phys.* **21**, 669 (1979).

[33] H.-O. Peitgen, H. Jürgens, and S. Saupe, *Chaos and Fractals: New Frontiers of Science* (New York: Springer-Verlag, 1992).

[34] Ibid., Chap. 12.

[35] E.N. Lorenz, "Predictability: Does the Flap of a Butterfly's Wings in Brazil Set off a Tornado in Texas," American Assoc. for the Advancement of Science, Meeting in Wash., DC (Dec. 1972).

[36] Hilborn, op. cit., Chap. 8.

[37] Kolmogorov-Arnol'd-Moser tori, discussed in V.I. Arnol'd, *Mathematical Methods in Classical Mechanics* (New York: Springer-Verlag, 1978).

[38] T. Shinbrot and F.J. Muzzio, *Phys. Today* **53**, 25 (Mar. 2000).

[39] E.B. Norman et al., *Phys. Rev. Lett* **60**, 2246 (1988).

[40] A. Petridis et al., *Bull. Am. Phys. Soc.* **47**, 97 (2002).

[41] A.I. Baz', Ya.B. Zel'dovich, and A.M. Perelomov, *Rasseyanie, reaktsii i raspaldy v nerelyativistskoi kvantovoi mekhanike*, 2$^{nd}$ Ed., (Moscow: Izdatel'stvo Nauka, 1971).

[42] A. Bohr and B.R. Mottelson, *Nuclear Structure*, Vol. 2 (Reading, MA: Benjamin Publ., 1975).

[43] Wm.C. McHarris, *J. Opt. B: Quantum and Semiclass. Opt.* **5**, S442 (2003).

[44] G.P. Williams, *Chaos Theory Tamed* (Washington DC: National Academy Press, 1997).

[45] J.S. Freedman and J.F. Clauser, *Phys. Rev. Lett.* **28**, 938 (1972); J.S. Clauser, *Phys. Rev. Lett.* **36**, 1223 (1976).

[46] A. Aspect et al., *At. Phys.* **8**, 103 (1983).

[47] W. Tittel et al., *Phys. Rev. Lett.* **81**, 3563 (1998).

[48] D.M. Greenberger, M.A. Horne, A. Shimony, and A. Zeilinger, *Am. J. Phys.* **58**, 1131 (1990); known as GHZ because of an preliminary paper that received numerous citations: D.M. Greenberger, M. Horne, and A. Zeilin-

ger, in *Bell's Theorem, Quantum Theory, and Conceptions of the Universe*, M. Kaftos, Ed., (Dordrecht: Kluwer Academic, 1989).

[49] N.D. Mermin, *Am. J. Phys.* **49**, 940 (1981); *Phys. Today* **38**, 38 (Apr. 1985).

[50] D.Z. Albert, *Quantum Mechanics and Experience* (Cambridge, MA: Harvard Univ. Press, 1992).

[51] K. Hess and W. Philipp, *Europhys. Lett.* **57**, 775 (2002); *Proc. Natl. Acad. (USA)* **98**, 14224 (2001); *Proc. Natl. Acad. (USA)* **98**, 14228 (2001); quant-ph/0103028; quant-ph/0206046 (2002); quant-ph/0305037; presented at the Växjö Conference, June 2003 (to appear in the Proceedings).

[52] N.D. Mermin, quant-ph/0206118 (2002).

[53] R.D. Gill, G. Weihs, A. Zeilinger, and M. Zukowski, quant-ph/0204269 (2002); D.M. Appleby, quant-ph/0210145 (2002); W.C. Myrvold, Univ. of Western Ontario, preprint (2003).

[54] A. Khrennikov, quant-ph/0205022 (2002); E.R. Loubenets, quant-ph/0309111 (2003).

[55] M. Schroeder, *Fractals, Chaos, and Power Laws* (New York: W.H. Freeman, 1991), Chap.4.

[56] S, Wolfram, *A New Kind of Science* (Champaign, IL: Wolfram Media, 2002).

[57] C. Tsallis, *J. Stat. Phys.* **52**, 479 (1988); E.M.F. Curado and C. Tsallis, *J. Phys. A: Math. GEN.* **24**, L69 (1991).

[58] C. Tsallis, presentation at DICE Workshop, Piombino, Italy, Sept. 2002; private communication (2002).

[59] Among the many papers published on the subject, the one perhaps most relevant to our subject is: V. Latora, M. Baranger, A. Rapisarda, and C. Tsallis, *Phys. Lett. A* **273**, 97 (2000).

[60] The most recent was the Workshop on "Anomalous Distributions, Nonlinear Dynamics and Nonextensivity," Santa Fe, NM, 6-9 Nov. 2002; proceedings to be published as *Nonextensive Entropy—Interdisciplinary Applications*, M. Gell-Mann and C. Tsallis, Eds. (2003).

[61] S. Aba, *Phys. Lett. A* **271**, 74 (2000).

[62] M.A. Nielsen and I.L. Chuang, *Quantum Computation and Quantum Information* (Cambridge: Cambridge Univ. Press, 2000).

[63] M.A. Nielsen, *Sci. Amer.* **287**, #5, 66 (2002).

[64] D. Tommasini, *J. High Ener. Phys.* **7**, 39 (2002)

[65] Wm.C. McHarris, *Analog* **CVI**, #1, 68 (1986).

[66] E. Schrödinger, *Naturwiss.* **23**, 807; **23**, 823; **23**, 844; J. Gribben, *In Search of Schrödinger's Cat* (New York: Bantam, 1984); *Schrödinger's Kittens and the Search for Reality* (Boston, MA: Little, Brown, 1995).

[67] R.P. Feynman, R.B. Leighton, and M. Sands, op. cit., Chaps. 1-3.

[68]  Wm.C. McHarris, *Proceedings of the Wigner Centennial Conference, Pécs, Hungary, July 2002,* Paper #24, published in on-line proceedings; also to be published in *Hungarian J. Phys.* (2003); Wm.C. McHarris, *Z. Naturforsch.* **a56**, 208 (2001).